\documentclass[pra,reprint,showpacs,aps,groupedaddress]{revtex4-2}

\usepackage{amsmath}
\usepackage{amssymb}
\usepackage{amsmath}
\usepackage{graphicx}
\usepackage[unicode]{hyperref}
\usepackage{bm}
\usepackage{ulem}
\usepackage{color}
\usepackage[T1]{fontenc}
\usepackage[utf8]{inputenc}
\usepackage{lmodern}
\usepackage{textcomp}
\usepackage{comment}
\usepackage{grffile}
\usepackage{algorithm}
\usepackage{algpseudocode}
\usepackage{physics}
\usepackage{natbib}

\newcommand{\hk}[1]{\textcolor[rgb]{0, 0, 0.0}{#1}}

\begin{document}

\title{Robust phase estimation of the ground-state energy without controlled time evolution on a quantum device}

\author{Hiroki Kuji$^{1,2}$}
    \email{1225702@ed.tus.ac.jp}
\author{Yuta Shingu$^{1}$}
\author{Tetsuro Nikuni$^{1}$}
\author{Takashi Imoto$^{3}$}
\author{Kenji Sugisaki$^{4,5,6,7}$}
\author{Yuichiro Matsuzaki$^{2}$}
    \email{ymatsuzaki872@g.chuo-u.ac.jp}

\affiliation{$^{1}$Department of Physics, Tokyo University of Science,1-3 Kagurazaka, Shinjuku, Tokyo, 162-8601, Japan}
\affiliation{$^{2}$Department of Electrical, Electronic, and Communication Engineering, Faculty of Science and Engineering, Chuo University}
\affiliation{$^{3}$Global Research and Development Center for Business by Quantum-AI Technology (G-QuAT), AIST, Central2, 1-1-1Umezono, Tsukuba, Ibaraki 305-8568, Japan}
\affiliation{$^{4}$Graduate School of Science and Technology, Keio University, 7-1 Shinkawasaki, Saiwai-ku, Kawasaki, Kanagawa 212-0032, Japan}
\affiliation{$^{5}$Quantum Computing Center, Keio University, 3-14-1 Hiyoshi, Kohoku-ku, Yokohama, Kanagawa 223-8522, Japan}
\affiliation{$^{6}$Keio University Sustainable Quantum Artificial Intelligence Center (KSQAIC), Keio University, 2-15-45 Mita, Minato-ku, Tokyo 108-8345, Japan}
\affiliation{$^{7}$Centre for Quantum Engineering, Research and Education, TCG Centres for Research and Education in Science and Technology, subspace V, Salt Lake, Kolkata 700091, India}

\date{\today}

\begin{abstract}
Estimating the ground-state energy of Hamiltonians in quantum systems is an important task. In this work, we demonstrate that the ground-state energy can be accurately estimated without controlled time evolution by using adiabatic state preparation (ASP) and Ramsey-type measurement. By considering the symmetry of the Hamiltonian governing the time evolution during ASP, we can prepare a superposition of the ground state and reference state whose eigenvalue is known. This enables the estimation of the ground-state energy via Ramsey-type measurement. Furthermore, our method is robust against non-adiabatic transitions, making it suitable for use with early fault-tolerant quantum computers and quantum annealing. 
\end{abstract}

\maketitle

\section{Introduction}\label{sec:introduction}
In quantum many-body systems and quantum chemistry, determining the ground-state energy is often important.
However, since the Hilbert space dimension of the Hamiltonian grows exponentially with the number of constituent elements (eg, the number of electrons), it becomes challenging to compute these eigenvalues by numerically diagonalizing the Hamiltonian  using classical computers.
To address this issue, several algorithms have been proposed to estimate energy eigenvalues of quantum many-body Hamiltonians using quantum computers, including those designed for noisy intermediate-scale quantum devices~\cite{Huggins2020-es,McClean2016-mn,O-Malley2016-en,Peruzzo2014-by,Parrish2019-mf,Stair2020-zl,Somma2019-lo,PRXQuantum.2.020317,PRXQuantum.2.020321}, early fault-tolerant quantum computer (FTQC)~\cite{Layden2022-xx,Campbell2022-we,Babbush2021-fj,Booth2021quantumaccelerated,PRXQuantum.3.010318}, and fully FTQC~\cite{Abrams1999-ok,Poulin2009-dr,Lin2020nearoptimalground,Ge2019-bc}.
One approach to estimate the eigenvalues of the Hamiltonian $\hat{H}$ is the phase estimation algorithm. In this method, an initial state is prepared by using adiabatic state preparation (ASP). The eigenvalues of the Hamiltonian are encoded in the phases of quantum states by applying a controlled version of the time evolution operator $\hat{U}=e^{-i\hat{H}t}$, which causes the quantum
state to accumulate a relative phase that corresponds to
the energy eigenvalue of the Hamiltonian. By measuring the phases
of the quantum states after evolution using the phase
estimation algorithm, one can extract the eigenvalues of
the Hamiltonian~\cite{Cleve1998-sw,Knill2007-bo,Nielsen2002-vo,Poulin2009-fg,Berry2009-xf,Higgins2007-yy,Kitaev2002-xd}. However, combining ASP with phase estimation involves controlled time-evolution operations, which can lead to prohibitively high implementation costs. To overcome this challenge, subsequent research proposed methods for estimating energy gaps without controlled time evolution~\cite{sugisaki2021bayesian,Sugisaki2020-po,sugisaki2023projective,matsuzaki2021direct,russo2021evaluating}. In particular, methods proposed in Refs.~\cite{matsuzaki2021direct,russo2021evaluating} use ASP and Ramsey-type measurements to estimate the energy differences. Although such methods can estimate the energy differences of the eigenvalues, directly estimating the eigenvalues remains challenging.

In this paper, we propose a method to estimate the ground-state energy directly, without requiring controlled time evolutions. Our approach leverages the symmetry of the Hamiltonian and utilizes trivial eigenstates, referred to as reference states, of the driving and problem Hamiltonians,which can be identified by a classical computer. The initial state is prepared as a superposition of the ground state and the reference state of the driving Hamiltonian $\hat{H}_{\mathrm{D}}$.
Starting from the above initial state and performing the ASP, we obtain the superposition of the ground state and the reference state of the problem Hamiltonian $\hat{H}_{\mathrm{P}}$. After evolving the system under the problem Hamiltonian for a period $\tau$, we apply the reverse ASP (RASP)~\cite{perdomo2011study,ohkuwa2018reverse,yamashiro2019dynamics,PhysRevResearch.3.033006} and project the state onto the initial state. By applying a Fourier transform of
the measurement results as a function of $\tau$, we extract the ground-state and the reference state of $\hat{H}_{\mathrm{P}}$. Since the energy
of the reference state is known, this energy difference allows us to determine the ground-state energy.
Furthermore, our detailed analysis demonstrates this method is robust against non-adiabatic transitions.

\section{Method}\label{sec:method}
\begin{figure*}[t]
    \centering
    \includegraphics[width=\textwidth]{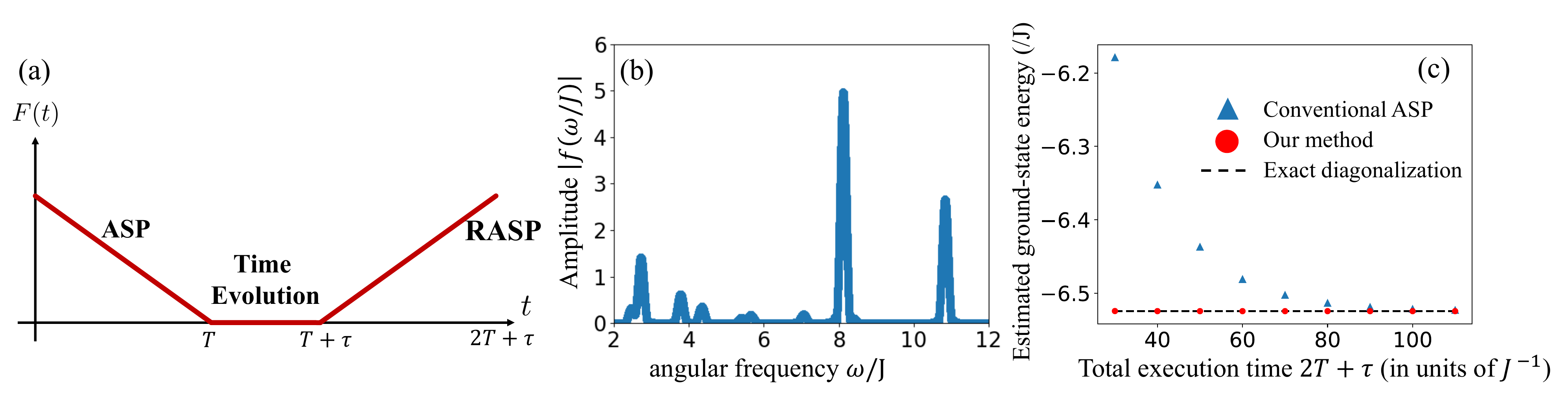}
    \caption{(a) A conceptual diagram of our method. The initial state is a superposition of the ground state of the driving Hamiltonian and a reference state of $\hat{H}_{\mathrm{D}}$. The system's Hamiltonian $\hat{H}(t)$ is expressed as a time-dependent combination of the driving and problem Hamiltonians, as shown in Eq~\eqref{eq:total_hamiltonian}, with the function $F(t)$ expressed in Eq.~\eqref{eq:F(t) scheduling}.
    For $0 \leq t \leq T$, the Hamiltonian corresponds to the standard ASP Hamiltonian, denoted as $\hat{H}_{\mathrm{ASP}}$. Subsequently, for $T \leq t \leq T+\tau$, the Hamiltonian is represented as $\hat{H}_{\mathrm{TE}}$, during which time evolution encodes the relative phase corresponding to the energy difference between the ground and the reference states of $\hat{H}_{\mathrm{P}}$ into the superposition state. Finally, for $T+\tau \leq t \leq 2T+\tau$, the Hamiltonian is denoted as $\hat{H}_{\mathrm{RASP}}$, which implements the RASP process. (b) The amplitude $|f(\omega/J)|$ obtained from the numerical simulation as a function of the frequency $\omega/J$. The peaks were observed at $\omega/J\simeq10.8, 8.1, 5.4, 3.8$, and $2.5$. These peaks correspond to the energy differences between the reference state of $\hat{H}_{\mathrm{P}}$ and the ground state, as well as the second, fifth, seventh, and tenth excited states, respectively. (c) Comparison of ground-state energy estimates obtained using conventional ASP and our proposed method. The horizontal axis represents the total runtime $2T+\tau$ ; for the conventional ASP, we set $T_{\mathrm{conv}}=2T+\tau$ to allow fair comparison.
    The vertical axis indicates the estimated ground-state energy. The blue markers represent the ground-state energy estimates obtained using the conventional method, while the red markers show the estimates calculated with our proposed method. The black dashed line represents exact diagonalization result.}
    \label{figure}
\end{figure*}
Here, we introduce our method for estimating the ground-state energy without controlled time evolution. The implementation of this method involves the following Hamiltonian:
\begin{align}
    \hat{H}(t)&=F(t)\hat{H}_{\mathrm{D}}+\qty[1-F(t)]\hat{H}_{\mathrm{P}},
    \label{eq:total_hamiltonian}\\
    F(t)&=\begin{cases}
        1-\frac{t}{T} & (0\leq t \leq T),\\
        0 & (T\leq t \leq T+\tau),\\
        \frac{t-(T+\tau)}{T} & (T+\tau \leq t \leq2T+\tau),
    \end{cases}
    \label{eq:F(t) scheduling}
\end{align}
where $F(t)$ is an external control parameter as shown in Fig~\ref{figure}~(a), $\hat{H}_{\mathrm{D}}$ and $\hat{H}_{\mathrm{P}}$ are the driving Hamiltonian and the problem Hamiltonian, respectively. We begin by describing a simplified scenario where the dynamics are adiabatic, although we will later address the effect of non-adiabatic transitions. The time-dependent Hamiltonian in Eq.~\eqref{eq:total_hamiltonian} takes the following three forms depending on the time region:
\begin{align}
        \hat{H}_{\mathrm{ASP}} = \qty(1-\frac{t}{T}) \hat{H}_{\mathrm D} + \frac{t}{T} \hat{H}_{\mathrm P}\quad (0\leq t \leq T)&\\
        \hat{H}_{\mathrm{TE}} = \hat{H}_{\mathrm P} \quad (T\leq t \leq T+\tau)&\\
        \notag\hat{H}_{\mathrm{RASP}} =  \frac{t-(T+\tau)}{T}\hat{H}_{\mathrm D} + \qty[1 - \frac{t-(T+\tau)}{T}] \hat{H}_{\mathrm P}\\ \quad (T+\tau \leq t \leq 2T+\tau)&.
\end{align}
For $0 \leq t \leq T$, the Hamiltonian is the standard ASP Hamiltonian $\hat{H}_{\mathrm{ASP}}$. During the next interval, $T \leq t \leq T+\tau$, the state evolves under the problem Hamiltonian $\hat{H}_{\mathrm{P}}$, and the superposition of the ground state and the reference state of $\hat{H}_{\mathrm{P}}$ acquires a relative phase corresponding to the energy gap between the two states, which is similar to the Ramsey-type measurement. Finally, during $T+\tau \leq t \leq 2T+\tau$, the system undergoes RASP under the Hamiltonian $\hat{H}_{\mathrm{RASP}}$~\cite{perdomo2011study,ohkuwa2018reverse,yamashiro2019dynamics,PhysRevResearch.3.033006}.

In our method, we make the following assumptions:
\begin{itemize}
    \item The driving Hamiltonian and the problem Hamiltonian each have a common conserved quantity $\hat{Q}$, satisfying $\comm*{\hat{H}_{\mathrm{D}}}{\hat{Q}}=\comm*{\hat{H}_{\mathrm{P}}}{\hat{Q}}=0$.
    \item  The operator $\hat{Q}$ is diagonalizable by a classical computer, and the number of distinct eigenvalues is $O(\mathrm{poly}(N))$, where $N$ is the number of qubits in the system under consideration.
    \item There exist degenerate eigenstates of $\hat{Q}$ with an eigenvalue of $q$ such that the number degeneracies is $O(\mathrm{poly}(N))$.
\end{itemize}

For example, quantities such as the total magnetization $\hat{M}  = \sum_{i=1}^{N} \hat{\sigma}^z_i$ can be used as $\hat{Q}$, where $N$ is the number of qubits and $\hat{\sigma}^z_i$ denotes the $z$-component of the Pauli operator for the $i$-th qubit. When these assumptions are satisfied,
$\hat{H}_{\mathrm{D}}$ and $\hat{H}_{\mathrm{P}}$ are block-diagonalized for each value of $q$. 
It is important to note that one of the subspaces defined by the eigenstates of $\hat{Q}$ contains the global ground state of the problem Hamiltonian.
For a given eigenvalue $q$ as stated above, the corresponding subspace is of size 
$O(\mathrm{poly}(N))$, making it diagonalizable using a classical computer. The eigenstates obtained in that subspace are denoted as $\ket*{\psi_{q,\mathrm{D}}}_{n}$ and $\ket*{\psi_{q,\mathrm{P}}}_{n}$, where $n$ represents the energy level within the subspace associated with the eigenvalue $q$.
We refer to these eigenstates as “reference states” in our proposed method. Note that our approach does not require all subspaces to be of size $O(\mathrm{poly}(N))$); we only require this property for those subspaces from which we choose our reference states. We assume that the energy of the reference state $\ket*{\psi_{q,\mathrm{P}}}_{n}$ has the largest energy in the problem Hamiltonian.
If the given problem Hamiltonian does not naturally satisfy this assumption, we can add an additional term such as $H_{\rm{add}}=-\lambda (\hat{Q}-q)^2$ where $\lambda$ is a positive hyperparameter.
It is worth noting that the eigenvalues of these reference states can be determined using a classical computer.

The use of a trivial eigenstate and its superposition with the ground state for estimating the ground-state energy of the quantum chemistry Hamiltonian via ASP was briefly mentioned in~\cite{sugisaki2021quantum} but no detailed explanation was provided.
On the other hand, our proposed method provides a much more general way to create and use the reference state for the ground-state energy estimation without control time evolution.

Using a superposition of the reference state of $\hat{H}_{\mathrm{D}}$ in the subspace of a given eigenvalue $q$ and the ground state of $\hat{H}_{\mathrm{D}}$ in the subspace of a given eigenvalue $q'$ as the initial state, we have 
\begin{align}
    \ket*{\psi_{\mathrm{ini}}}=\frac{1}{\sqrt{2}}(\ket*{\psi^{\mathrm g}_{q',\mathrm{D}}}+\ket*{\psi_{q,\mathrm{D}}}_{n}),
\end{align}
where $\ket*{\psi^{\mathrm{g}}_{q',\mathrm{D}}}$ is the ground state of the driving Hamiltonian in the subspace of a given eigenvalue $q'$. After ASP under $\hat{H}(t)$, the system evolves into: 
\begin{align}
    \ket*{\psi}=\frac{1}{\sqrt{2}}(\ket*{\psi^{\mathrm g}_{q',\mathrm{P}}}+e^{i\theta}\ket*{\psi_{q,\mathrm{P}}}_{n}),
\end{align}
where $\ket*{\psi^{\mathrm g}_{q',\mathrm{P}}}$ is the ground state of the problem Hamiltonian in the subspace of a given eigenvalue $q'$, and $\theta$ is the relative phase arising from the ASP process. Here, $\ket*{\psi^{\mathrm{g}}_{q',\mathrm{P}}}$ is the state obtained by performing ASP, which is associated with $\ket*{\psi^{\mathrm{g}}_{q',\mathrm{D}}}$. Then, the state evolves under $\hat{H}_{\mathrm {P}}$ for a time $\tau$, leading to 
\begin{align}
    e^{-i\hat{H}_{\mathrm P}\tau}\ket*{\psi}=\frac{1}{\sqrt{2}}(e^{-iE^{\mathrm g}_{q',\mathrm P}\tau}\ket*{\psi^{\mathrm g}_{q',\mathrm{P}}}+e^{-iE^{n}_{q,\mathrm{P}}\tau+i\theta}\ket*{\psi_{q,\mathrm{P}}}_{n}),
\end{align}
where $E^{n}_{q,\mathrm{P}}$ represents the eigenvalue of $\ket*{\psi_{q,\mathrm{P}}}_{n}$, which we can obtain using a classical calculator.
Subsequently, RASP is executed under $\hat{H}_{\mathrm{RASP}}$, resulting in the final state: 
\begin{align}
    \ket*{\psi_{\mathrm{f}}}=\frac{1}{\sqrt{2}}(e^{-iE^{\mathrm{g}}_{q',\mathrm P}\tau}\ket*{\psi^{\mathrm{g}}_{q',\mathrm{D}}}
+e^{-iE^{n}_{q,\mathrm{P}}\tau+i\theta'}\ket*{\psi_{q,\mathrm{P}}}_{n}),
\end{align}
where $\theta'$ is a relative phase, and $E^{\mathrm{g}}_{q',\mathrm P}$ is the ground-state energy of $\hat{H}_{\mathrm{P}}$. 
Note that any relative phase acquired during the ASP process does not affect the final result of the eigenvalue estimation (see Appendix~\ref{append: relative phase} for details). 
The above sequence is summarized in Fig.~\ref{figure}~(a). A measurement is performed using the projection operator $\hat{\mathcal{M}} = \ketbra{\psi_{\mathrm{ini}}}$, and the probability of projecting onto this state is given by: 
\begin{align}
    \mathcal{P}(\tau)\equiv\abs{\ip{\psi_{\mathrm{f}}}{\psi_{\mathrm{ini}}}}^2=\cos^2\qty[\frac{(E^{\mathrm g}_{q',\mathrm P}-E^{n}_{q,\mathrm{P}})\tau}{2}+\frac{\theta'}{2}].
    \label{eq:projection}
\end{align}
This signal oscillates with the frequency corresponding to the energy difference. By repeating the above steps and sweeping $\tau$, multiple measurement results are obtained. Finally, a discrete Fourier transform of $\mathcal{P}(\tau)$ is performed as 
\begin{align}
    f(\omega)=\sum_{n=1}^{L}\mathcal{P}(\tau_n)e^{-i\omega \tau_n},
    \label{eq:discrete fourier transformation}
\end{align}
where the discretized time step $\tau_n$ is defined as $\tau_n = \tau_{\mathrm{min}} + \frac{n-1}{L-1}(\tau_{\mathrm{max}} - \tau_{\mathrm{min}})$, with $\tau_{\mathrm{min}}$ and $\tau_{\mathrm{max}}$ denoting the minimum and maximum times, and $L$ is the number of steps. 
\hk{Let us denote the uncertainty of the estimated energy as $\delta \omega$. According to the time–energy uncertainty relation, $1/\Delta\tau \simeq \delta \omega$, where $\Delta \tau=(\tau_{\mathrm{max}}-\tau_{\mathrm{min}})/(L-1)$ is the time interval, which can be chosen to match the desired estimation precision.}

The peak of $f(\omega)$ appears at $\omega$ corresponding to the energy difference $E^{\mathrm{g}}_{q',\mathrm{P}} - E^{n}_{q,\mathrm{P}}$. Since $E^{n}_{q,\mathrm{P}}$ is known, the ground-state energy in the subspace corresponding to eigenvalue $q'$ can be determined. From this procedure, we successfully estimate the ground-state energy in the subspace with $q'$. To estimate the ground-state energy in another subspace with $q''$, we could perform the above procedure, say $q''$. Thanks to the second assumption of our protocol, we can perform this search across all subspaces within polynomial time. Similarly, the first-excited-state energy in the subspace of $q'$ can be estimated by preparing a superposition of the first excited state and the reference state of $\hat{H}_{\mathrm{D}}$ in the subspace of eigenvalues $q'$ and $q$ respectively, as the initial state.

Furthermore, if non-adiabatic transitions occur during the ASP, peaks appear in $f(\omega)$ at frequencies corresponding to the energy difference between the excited state and the reference state of $\hat{H}_{\mathrm{P}}$, such as $E^{\mathrm{e_1}}_{q',\mathrm {P}}-E^{n}_{q,\mathrm{P}},E^{\mathrm{e_2}}_{q',\mathrm {P}}-E^{n}_{q,\mathrm{P}},\dots$, where $E^{\mathrm{e}_i}_{q',\mathrm {P}}$ denotes the $i$-th excited-state energy in the subspace associated with eigenvalue $q'$. 
These excitation energies can also be estimated since the trivial eigenvalue is known in advance. 
\hk{This implies that our method does not require careful tuning of the ASP execution time $T$ in advance. This is in stark contrast to the conventional scheme, where the ground state is prepared adiabatically and the ground-state energy is estimated by measuring the expectation value of the Hamiltonian. In such conventional methods, even a small non-adiabatic transition can significantly reduce the fidelity of the prepared state with respect to the true ground state of the problem Hamiltonian, resulting in inaccurate energy estimation.}

\hk{Specifically, the problem Hamiltonian can be decomposed as $\hat{H}_{\mathrm P}=\sum_j c_j \hat{P}_j$, where $c_j$ is a real number and $\hat{P}_j$ is a tensor product of the Pauli operators.
In conventional methods, the ground-state energy is estimated by preparing the ground state via ASP and then measuring the expectation value $\langle \hat{H}_{\mathrm P}\rangle =\sum_j c_j \langle \hat{P}_j \rangle $. To ensure high fidelity with the true ground state, the total execution time $T_{\rm conv}$ must satisfy the adiabatic condition~\cite{childs2001robustness,morita2008mathematical,albash2018adiabatic,hauke2020perspectives}. 
The time-dependent Hamiltonian in the conventional ASP method is typically given as $\hat{H}_{\rm conv}=(1-s)\hat{H}_{\mathrm D}+s\hat{H}_{\mathrm P}$, where $s=t/T_{\rm conv}$ the normalized time, $t$ is the physical time, and 
$T_{\rm conv}$ is the total execution time of ASP. 
The adiabatic condition is then explained as
\begin{align}
     \max_{s \in [0,1]}\frac{|\bra{\psi^{\mathrm{e}_n}(s)}\dot{\hat{H}}(s)\ket{\psi^{\mathrm{g}}(s)}|}{{|E^{\mathrm{e}_n}(s)-E^{\mathrm{g}}(s)|}^{2}}\ll T_{\rm conv} ,\label{eq:adiabatic_criterion}
\end{align}
where  $\ket{\psi^{\mathrm{e}_n}(s)}$ and $\ket{\psi^{\rm g}(s)}$ denote the $n$-th excited and ground state of the instantaneous Hamiltonian $\hat{H}(s)$, respectively, and $E^{\mathrm{e}_n}(s)$ and $E^{\rm g}(s)$ are their corresponding eigenenergies. $\dot{\hat{H}}(s)$ is the derivative of $\hat{H}(s)$ with respect to the normalized time $s$. In practice, neither the transition matrix elements nor the energy gaps of the Hamiltonian are known in advance, making it difficult to determine a suitable value of $T_{\rm conv}$ that ensures a high fidelity beforehand. In contrast, our method allows for accurate energy estimation even in the presence of non-adiabatic transitions, by utilizing interference between the ground state and a reference state. Therefore, our approach eliminates the need for fine-tuning of the ASP runtime and provides a practical
advantage over the conventional expectation-value-based method. It should be noted, however, if non-adiabatic transitions significantly reduce the ground state population, both our method and the conventional quantum phase estimation fail to determine the ground-state energy.}

Regarding the reference state, since it is chosen from a subspace of size $O(\mathrm{poly}(N))$, non-adiabatic transitions can be suppressed byu setting $T$ appropriately within polynomial time. Therefore, non-adiabatic transitions involving the reference state are considered negligible in our analysis.

Finally, our scheme is particularly effective for both early FTQC~\cite{Suzuki2022-lc} and quantum annealing~\cite{kadowaki1998quantum,farhi2000quantum,imoto2022quantum,farhi2001quantum} since it does not require controlled time-evolution, which leads to shorter quantum circuits.

\section{Result}\label{sec:result}
We estimate the ground-state energy of the spin$-1/2$ Heisenberg model using our proposed method through numerical simulations.
In this study, we consider a 4-qubit system. The problem Hamiltonian is defined as the Heisenberg model with an additional inhomogeneous longitudinal magnetic field to lift the degeneracy:
\begin{align}
    \hat{H}_{\mathrm P}=J\sum_{i=1}^N(\hat{\sigma}^x_i\hat{\sigma}^x_{i+1}+\hat{\sigma}^y_i\hat{\sigma}^y_{i+1}+\hat{\sigma}^z_i\hat{\sigma}^z_{i+1})+\sum_{i=1}^N B'_i\hat{\sigma}^z_i,
    \label{eq:problem hamiltonian}
\end{align}
where $J$ is the exchange interaction strength, and $B'_i$ represents the magnetic field acting on the $i$-th qubit. The exchange interaction strength is set to $J=1$, and the magnetic field strengths are set to $B'_1/J=-0.24, B'_2/J=-0.34, B'_3/J=-0.62$, and $B'_4/J=-0.09$.
Periodic boundary conditions are applied.
To facilitate the preparation of the ground state, we introduce the following driving Hamiltonian:
\begin{align} 
    \hat{H}_{\mathrm D}=\sum_{i=1}^{N/2} J_{2i-1,2i}(\hat{\sigma}^x_{2i-1}\hat{\sigma}^x_{2i}+\hat{\sigma}^y_{2i-1}\hat{\sigma}^y_{2i})+\sum_{i=1}^N B_i\hat{\sigma}^z_i,
    \label{eq: driving Hamiltonian}
\end{align}
where $J_{2i-1,2i}$ is the interaction strength between qubits, $B_i$ denotes the strength of the magnetic field acting on the $i$-th qubit. 
The parameters are set as $J_{1,2}/J=0.5$, $J_{3,4}/J=0.3$, $B_{1}/J=B_2/J=-1$, and $B_{3}/J=B_4/J=1$. Under these settings, the ground state of the driving Hamiltonian is $\ket*{1100}$.
Here, we consider the case where $\hat{Q} = \hat{M}$, which clearly commutes with both $\hat{H}_{\mathrm{D}}$ and $\hat{H}_{\mathrm{P}}$. Since the states $\ket*{11\cdots 1}$ and $\ket*{00\cdots 0}$ are the unique eigenstates of $\hat{M}$ corresponding to the eigenvalues $\pm N$, they are trivial simultaneous eigenstates of both $\hat{H}_{\mathrm{D}}$ and $\hat{H}_{\mathrm{P}}$, with their corresponding trivial eigenvalues denoted as $E_{\pm N,\mathrm{D}}$ and $E_{\pm N,\mathrm{P}}$, respectively.
The initial state is prepared as follows: $\ket*{\Psi_0}=\frac{1}{\sqrt{2}}(\ket*{1100}+\ket*{1111})$ (see Appendix~\ref{append: initial state prep} for details).
Our proposed method is implemented starting from this state.

Figure~\ref{figure}~(b) shows the Fourier transform $\abs{f(\omega/J)}$ plotted against $\omega/J$. We set the execution time for ASP as $JT=5$ and varied $\tau$ from $J\tau_{\mathrm{min}}=0$ to $J\tau_{\mathrm{max}}=70$ in $L=1000$ steps. The peaks observed at $\omega/J\simeq 10.8, 8.1, 5.4, 3.8, 2.5$ correspond to the energy differences between the estimated energy eigenvalues and the trivial energy eigenvalue. In the case of the Heisenberg model, for the above parameters, the reference state $\ket*{1111}$ corresponds to the maximum energy eigenvalue. 
Due to this, we can identify the ground-state energy from the largest energy difference.
Specifically, the peak at $\omega/J=10.8$ corresponds to the energy difference between the trivial eigenvalue $E_{N,\mathrm{P}}$ and the ground-state energy $E^{\mathrm g}_{0,\mathrm{P}}$ of the problem Hamiltonian. Given that $E_{N,\mathrm{P}}/J=4.3$, the ground-state energy is estimated as $E^{\mathrm g}_{0,\mathrm{P}}/J\simeq-6.524590$.
In comparison, the ground-state energy obtained using numerical diagonalization is $E^{\mathrm g}_{0,\mathrm{P}}/J\simeq-6.524593$, corresponding to a relative error of $3.7\times 10^
{-5}\%$.
Furthermore, the energy levels of the excited states obtained via numerical diagonalization, all expressed in units of $J$, are approximately $-3.80585, -1.09182, 0.53380$ and $1.84517$ for the second, fifth, seventh, and tenth excited states, respectively. The corresponding values estimated from the peaks of $f(\omega)$ are around $-3.80588$, $-1.08495$, $0.53388$ , and $1.84942$, respectively. The relative errors between the exact and estimated values are $6.0\times10^{-4}\%$, $6.0\times10^{-1}\%$, $1.4\times 10^{-2}\%$, and $2.3\times 10^{-1}\%$, respectively.

There is a tendency for the relative error of the estimation to increase as the eigenenergy of the excited state increases. A possible explanation is as follows. The trivial energy eigenvalue is the largest among the eigenenergies of the problem Hamiltonian. Consequently, the energy gap between the reference state and higher excited states becomes smaller. To estimate this small energy gap, a longer time $\tau$ is required for the Ramsey-type measurement due to the time-energy uncertainty relation.
To achieve high accuracy in estimating the excitation energies, the initial state should be prepared as a superposition of the reference state and the excited state of the driving Hamiltonian, and then the proposed method can be applied. Alternatively, changing the reference state or choosing a sufficiently long $\tau$ can also  overcome the challenge of accurately estimating the eigenenergy of the excited state.

Furthermore, we use the conventional ASP method to calculate the ground-state energy, which is obtained from the expectation value after the ASP process, and compare it with the ground-state energy values estimated by our proposed method. The results shown in Fig.~\ref{figure}~(c) clearly demonstrate that our proposed method can estimate the ground-state energy with high accuracy in a shorter execution time compared to the conventional ASP method. Additionally, while the conventional method fails to accurately estimate the ground-state energy at shorter execution times due to the effects of non-adiabatic transitions, our proposed method exhibits robustness against these effects, consistently providing accurate estimations.

In quantum chemistry calculations, a situation where electrons occupy all orbitals, the reference state is known to be $\ket*{1 \dots 1}$ under Jordan--Wigner transformation, which in general corresponds to the maximum energy eigenstate. In this way, the ground-state energy can be calculated using our method. 

\section{Conclusion}\label{sec:conclusion}
In this study, we propose a method to accurately estimate the ground-state energy of a Hamiltonian using ASP in early FTQC or quantum annealing, without requiring controlled time evolution. Our approach leverages the symmetry of the Hamiltonian, enabling the direct estimation of the ground-state energy by preparing an initial state that is a superposition of the ground state and a reference state. Unlike the conventional method~\cite{matsuzaki2021direct,russo2021evaluating} for estimating the energy gap without controlled time evolution, our method directly estimates the ground-state energy eigenvalue in the subspace of interest. \hk{To identify the true ground-state energy of the problem Hamiltonian from the full spectrum, a search across all subspaces can be performed in polynomial time, owing to the second assumption of our protocol (see Appendix~\ref{append: different subspace} for details). Furthermore, a comparison with the result obtained by the conventional ASP reveals that our method offers a clear practical advantage. It is robust against non-adiabatic transitions, eliminating the need for fine-tuning the execution time $T$ and enabling accurate ground-state energy estimation even with short execution times. This is in stark contrast to the conventional ASP, which relies on strict adiabaticity. Moreover, it can estimate the energies of excited states induced by non-adiabatic transitions.} 
Future work includes further experimental verification of this method, exploration of its applicability to other quantum systems, and generalization of the symmetry utilized in the initial state preparation.

\section{Acknowledgment}
We thank Suguru Endo for his useful comments. This work is supported by JSPS KAKENHI (Grant Number 23H04390, 20H05661), JST Moonshot (Grant Number JPMJMS226C), CREST (JPMJCR23I5), and presto JST (JPMJPR245B). K.S. acknowledges support from Quantum Leap Flagship Program (Grant No. JPMXS0120319794) from the MEXT, Japan, Center of Innovations for Sustainable Quantum AI (JPMJPF2221) from JST, Japan, and Grants-in-Aid for Scientific Research C (21K03407) and for Transformative Research Area B (23H03819). 
\appendix
\clearpage
\begin{widetext}
\section{Effect of relative phases on phase estimation}\label{append: relative phase}
In this appendix, we provide a mathematical proof that the relative phase factor $e^{i\theta}$ of the superposition state $\ket{\psi}$ between different subspaces used in our method does not affect the final eigenvalue estimation, regardless of its specific value.

Let us rewrite $\mathcal{P}(\tau)$ given by Eq.~\eqref{eq:projection} as
\begin{align}
    \mathcal{P}(\tau)=\cos^2\qty(\frac{\Delta E\tau}{2}+\frac{\theta'}{2})=\frac{1}{2}+\frac{1}{2}\cos\qty(\frac{\Delta E\tau}{2}+\frac{\theta'}{2}),
\end{align}
where we define $\Delta E\equiv E^{\mathrm g}_{q',\mathrm P}-E^{n}_{q,\mathrm{P}}$.
Substituting this into Eq.~\eqref{eq:discrete fourier transformation}, we obtain
\begin{align}
    f(\omega)=\frac{1}{2} \sum_{n=1}^{L} e^{-i \omega \tau_{n}}+\frac{1}{2}\sum_{n=1}^{L}\cos\qty(\frac{\Delta E\tau_n}{2}+\frac{\theta'}{2}) e^{-i \omega \tau_{n}}.
    \label{eq:rewrited fourier}
\end{align}
Focusing on the second term on the right-hand side of Eq.~\eqref{eq:rewrited fourier}, where the relative phase $\theta'$ appears, we use the identity $\cos(\phi)=\Re(e^{i\phi})$ to express it as
\begin{align}
    \notag \frac{1}{2}\sum_{n=1}^{L}\cos\qty(\frac{\Delta E\tau_n}{2}+\frac{\theta'}{2}) e^{-i \omega \tau_{n}}
    &=\frac{1}{2}\Re\qty(\sum_{n=1}^{L}\exp\qty[i\qty(\frac{\Delta E\tau_n}{2}+\frac{\theta'}{2})] e^{-i \omega \tau_{n}})\\
    &=\frac{1}{2}\Re\qty(\exp\qty(\frac{i\theta'}{2})\sum_{n=1}^{L}\exp\qty[i\frac{(\Delta E-2\omega)\tau_n}{2}]).
\end{align}
The key observation here is that the relative phase $\theta'$ factors out as a global phase term. The Fourier transform determines the peak positions and magnitudes based on the term $\exp\qty[i\frac{(\Delta E-2\omega)\tau_n}{2}]$, while the phase factor $\exp\qty(i\frac{\theta'}{2})$ has an absolute value of $1$ and thus does not affect the final eigenvalue estimation.

\section{Initial state preparation via ASP}\label{append: initial state prep}
An implicit assumption of our proposal, as described in the main text, is that the initial state can be prepared as a superposition of the ground state (belonging to a subspace characterized by a specific eigenvalue $q$) of the driving Hamiltonian and the reference state. In this appendix, we discuss how such a state in the case of $\hat{Q}=\hat{M}$ can be prepared.
The driving Hamiltonian used in this study is given by Eq.~\eqref{eq: driving Hamiltonian}. The initial state is defined as 
\begin{align}
    \ket{\psi_{\mathrm{ini}}}=\frac{1}{\sqrt{2}}\qty(\ket{\psi^{\mathrm{g}}_{m,\mathrm{D}}}+\ket{11\cdots 1}),
    \label{eq.initial state}
\end{align}
where $\ket{\psi^{\mathrm{g}}_{m,\mathrm{D}}}$ represents the ground state of the driving Hamiltonian within the subspace corresponding to each eigenvalue $m$ of the conserved quantity $\hat{M}$, and $\ket{11\cdots 1}$ represents the reference state.

To begin with, the driving Hamiltonian in Eq.~\eqref{eq: driving Hamiltonian} should be diagonalized to identify the energy levels of $\ket{\psi^{\mathrm{g}}_{m,\mathrm{D}}}$ and the reference state $\ket{11\cdots1}$ within the full energy spectrum. Specifically, if diagonalization reveals that these states correspond to the $n$-th and $l$-th excited states,respectively, we can use this information to generate the desired initial state by using ASP. 

The basic strategy is as follows. The desired state is $\frac{1}{\sqrt{2}}(|\tilde{E}_{n,\mathrm{D}}\rangle +|\tilde{E}_{l,\mathrm{D}}\rangle )$ where $|\tilde{E}_{n,\mathrm{D}}\rangle $ ($|\tilde{E}_{l,\mathrm{D}}\rangle $) is the $n$-th ($l$-th) eigenstate of the driving Hamiltonian in Eq.~\eqref{eq: driving Hamiltonian} Once we create $\frac{1}{\sqrt{2}}(|\tilde{E}_{n}\rangle +|\tilde{E}_{l}\rangle )$ where $|\tilde{E}_{n}\rangle $ ($|\tilde{E}_{l}\rangle $) is the $n$-th ($l$-th) eigenstate of the Hamiltonian of the transverse magnetic field $\hat{H}_{\mathrm{transverse}}$, we can adiabatically prepare the desired state by changing the Hamiltonian from $\hat{H}_{\mathrm{transverse}}$ to $\hat{H}_{\mathrm{D}}$. Although it is known that we can prepare the so-called GHZ state, 
$\frac{1}{\sqrt{2}}(|\tilde{E}_{n}\rangle +|\tilde{E}_{l}\rangle )$ is slightly different from the GHZ state. Fortunately, 
after creating a small GHZ state, we can prepare 
just by adding separable qubits, $\frac{1}{\sqrt{2}}(|\tilde{E}_{n}\rangle +|\tilde{E}_{l}\rangle )$. By combining these strategy, our proposed prescription is as follows.

For simplicity, let us assume that the number of qubits are even. 
The procedure consists of the following steps:
\begin{enumerate}
    \item The process starts from the all-plus state $\ket{++\cdots +}$. We use ASP to transform this state into a GHZ state $|\psi _{\rm{GHZ}}\rangle =(\ket{00\cdots 0}+\ket{11\cdots1})\sqrt{2}$ by gradually switching the Hamiltonian from $\hat{H}_1=B_1\sum_{i}\hat{\sigma}^x_i$ to $\hat{H}_2=-B_2{(\sum_{i}\hat{\sigma^z_i})}^2$, where $B_1$ ($B_2$) represent the strength of the magnetic field (interaction between qubits)~\cite{choi2017quantum,lanting2014entanglement,lee2006adiabatic,yukawa2018fast,xing2016heisenberg,matsuzaki2022generation,hatomura2022quantum}.
    \item The Hamiltonian is then switched from $\hat{H}_2$ to $\hat{H}_3=-B_3{(\sum_{i}\hat{\sigma^x_i})}^2$ by using ASP, where $B_3$ represent the strength of the interaction between qubits, so that the state becomes $|\psi' _{\rm{GHZ}}\rangle =(\ket{++\cdots +}+\ket{--\cdots-})\sqrt{2}$. Before and after this ASP step, the system remains a superposition of degenerate ground states.
    Both states $|\psi _{\rm{GHZ}}\rangle$ and $|\psi'_{\rm{GHZ}}\rangle$ are eigenstates of a parity operator defined as $\hat{P}=\hat{\sigma}^x_1\otimes\hat{\sigma}^x_2\otimes\cdots\otimes\hat{\sigma}^x_N$.
    \item The Hamiltonian is switched from $\hat{H}_3$ to $\hat{H}_4=\sum_i B_i\hat{\sigma}^x_i$. Note that, because $\hat{H}_3$ and $\hat{H}_4$ commute, the state remains unchanged.
    \item By adding qubits to match the desired superposition, we can create a state such as $\ket{\pm \pm \cdots \pm} (\ket{++\cdots +}+\ket{--\cdots-})/\sqrt{2}$. After relabeling the qubits, this corresponds to $\frac{1}{\sqrt{2}}(|\Tilde{E}_n\rangle +|\Tilde{E}_l\rangle )$ where $|\Tilde{E}_n\rangle $ ($|\Tilde{E}_l\rangle $) is the $n$-th ($l$-th) eigenstate of $\hat{H}_{\mathrm{transverse}}=\sum_i B_i\hat{\sigma}^x_i$, where $\ket{\pm \pm \cdots \pm }$ represents a product state in which each qubit is in either $\ket{+}$ or $\ket{-}$.
    \item Finally, the Hamiltonian is switched from $\hat{H}_{\mathrm{transverse}}$ to $\hat{H}_{\mathrm D}$ (Eq.~\eqref{eq: driving Hamiltonian}) by using ASP. Then, we obtain the desired superposition state (Eq.~\eqref{eq.initial state}).
\end{enumerate}
At the final step, $\hat{H}_{\mathrm{transverse}}$ is the Hamiltonian without interaction while $\hat{H}_{\mathrm D}$ is the Hamiltonian with interaction only between the $2i-1$-th qubit and the $2i$-th qubit. This means that the evolution at the final step can be reduced to two-qubit dynamics. 
It is worth mentioning that we can reduce the number of qubits by measuring one of the qubits of the GHZ states with an even number of qubits if we need a GHZ state with an odd number of qubits.

\section{Phase estimation in different subspaces}\label{append: different subspace}
\begin{figure*}[h]
    \centering
    \includegraphics[width=\textwidth]{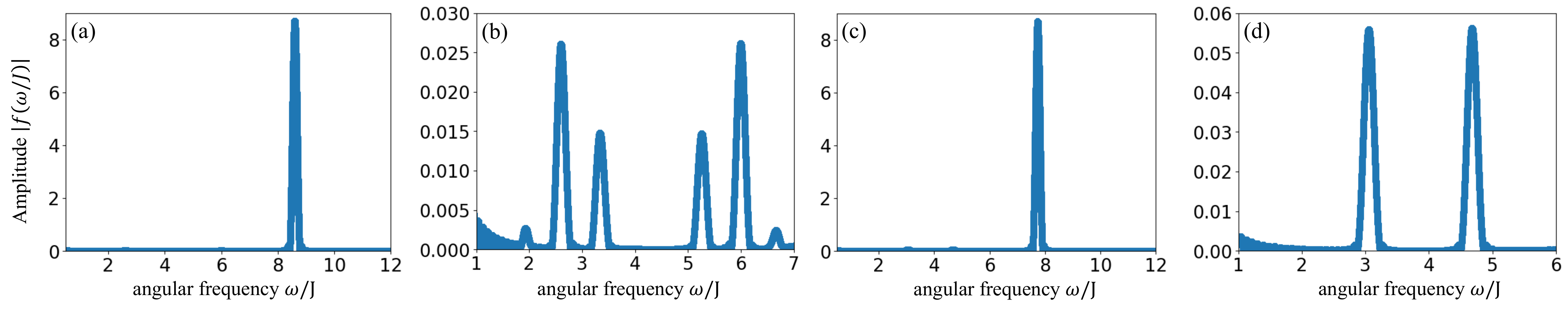}
    \caption{The amplitude $|f(\omega/J)|$ obtained from the numerical simulation as a function of the frequency $\omega/J$. (a) Results for a $m=2$ subspace. (b) A magnified view of a specific region in (a). (c) Results for a $m=-2$ subspace. (d) A magnified view of a specific region in (c). (a),(b) The peaks were observed at $\omega/J\simeq8.6, 6.0, 3.3$, and $1.9$. These peaks correspond to the energy differences between the reference state of $\hat{H}_{\mathrm{P}}$ and the first, fourth, eighth, and eleventh excited states, respectively. (c),(d) The peaks were observed at $\omega/J\simeq 7.7$ and $4.7$. These peaks correspond to the energy differences between the reference state of $\hat{H}_{\mathrm{P}}$ and the third, and sixth excited states, respectively. Note that the peaks not discussed here correspond to the energy differences between the excited states.}
    \label{fig:diff subspaces}
\end{figure*}

In this study, we propose a method for estimating the energy eigenvalues within a subspace characterized by a specific eigenvalue $q$ of a conserved quantity $\hat{Q}$, which commutes with both the driving Hamiltonian and the problem Hamiltonian. An key aspect of our method is that it does not require prior knowledge of which subspace contains the true ground state of the problem Hamiltonian. By applying our method to each subspace separately and comparing the obtained eigenvalues, we can identify the true ground state of the problem Hamiltonian. Of course, if the subspace containing the target eigenvalue of the problem Hamiltonian is known in advance through some other method, our approach can be directly applied to that subspace to obtain the desired eigenvalue. Alternatively, if the
ground state of a particular subspace is of interest, our method can also be used to identify it.

In Sec.~\ref{sec:result}, we considered the case of $\hat{Q}=\hat{M}$ and applied our proposed method to the subspace with eigenvalue $m=0$ to estimate the ground energy. 
In the main text, we made this choice
for simplicity
assuming that prior knowledge from exact diagonalization indicated that the true ground state of Eq.~\eqref{eq:problem hamiltonian} lies in the $m=0$ subspace. In this appendix, we remove the assumption. More specifically,
 we apply our method to the subspaces with $m=\pm2$ and verify that the true ground energy can be estimated from the obtained eigenvalues. However, we omit the cases of $m=\pm 4$, as these subspaces are both one-dimensional and therefore trivial.

Figure~\ref{fig:diff subspaces} (a) and (b) show the Fourier transform $\abs{f(\omega/J)}$ plotted against $\omega/J$ for $m=2$.
The numerical parameters used are the same as those in Sec.~\ref{sec:result}, except for the initial states. For the $m=2$ subspace, we set the initial state as a superposition of $\ket*{\psi^{\mathrm g}_{2,\mathrm{D}}}=\frac{1}{\sqrt{2}}(\ket{0010}-\ket{0001})$ and the reference state,
which was chosen as $\ket{1111}$. The results show peaks at $\omega/J\simeq 8.6, 6.0, 3.3$, and $1.9$. These peaks correspond to the energy differences between the estimated energy eigenvalues and the trivial energy eigenvalue. Given that $E_{4,\mathrm{P}}/J=4.3$, the estimated energy eigenvalues from the peaks of $\abs{f(\omega/J)}$ are approximately $-4.28497, -1.68560, 0.97414$, and $2.38026$, all expressed in units of $J$. The energy levels of the excited states obtained via numerical diagonalization, all expressed in units of $J$, are approximately $-4.28498, -1.68559, 0.97265$ and $2.37795$ for the first, fourth, eighth, and eleventh excited states, respectively.
Similarly, we plot the results for $m=-2$ in Fig.~\ref{fig:diff subspaces} (c) and (d). Here, the initial state is 
a superposition of $\ket*{\psi^{\mathrm g}_{-2,\mathrm{D}}}=\frac{1}{\sqrt{2}}(\ket{1011}-\ket{0111})$ and the reference state $\ket{1111}$.

The results in Fig.~\ref{fig:diff subspaces}(c) and (d) show peaks at $\omega/J\simeq 7.7$ and $4.7$. The estimated energy eigenvalues from the peaks of $\abs{f(\omega/J)}$ are approximately $-3.41947$ and $-0.36850$. The energy levels of the excited states obtained via numerical diagonalization, all expressed in units of $J$, are approximately $-3.41949$ and $-0.36855$ for the third and sixth excited states, respectively. Note that the peaks not discussed here correspond to the energy differences between the excited states. 
By comparing these results for $m=\pm 2$ with those for $m=0$, we can conclude that the energy $-6.524590$ mentioned in Sec.~\ref{sec:result} is indeed the true ground-state energy of the problem Hamiltonian.

As these results indicate, our proposed method does not require prior knowledge of which subspace contains the true ground energy of the problem Hamiltonian. By applying our method to different subspaces and comparing the obtained results, we can successfully estimate the true ground energy of the problem Hamiltonian.

\end{widetext}


\bibliographystyle{apsrev4-2} 
\bibliography{ref} 

\end{document}